\documentclass[10pt]{article}
\usepackage[OE]{express}
\usepackage{cite}
\begin{document}
\title{Generation of multiple vector beams from a single digital hologram}

\author{Carmelo Rosales-Guzm\'an \authormark{1,*} , Nkosiphile Bhebhe,\authormark{1}and  Andrew Forbes\authormark{1}}

\address{\authormark{1}School of Physics, University of the Witwatersrand, Johannesburg 2050, South Africa}
\email{\authormark{*}carmelo.rosalesguzman@wits.ac.za} 

\begin{abstract}
Complex vector light fields, classically entangled in polarization and phase, have become ubiquitous in a wide variety of research fields. This has triggered the demonstration of a wide variety of generation techniques. Of particular relevance are those based on computer-controlled devices due to their great flexibility. More specifically, spatial light modulators have demonstrated that almost any vector beam, with various spatial profiles and polarization distributions can be generated. Crucially, none of these techniques has proven capable to generate multiple vector beams simultaneously. Here, we put forward a novel technique that exploits the superposition principle in optics to enable the simultaneous generation of many vector beams using a single digital hologram. As proof-of-principle, we demonstrated the simultaneous generation of twenty vector vortex beams with various polarization distributions and spatial shapes on a single SLM, each of which with their own spatial shape and polarization distribution.  
\end{abstract}


\section{Introduction}
The ability to tailor the spatial properties of light has significantly changed the landscape of photonic-based applications. This is the case of complex light fields classically entangled in polarization and phase, a topic of late due to their various applications. In particular, classically entangled fields with cylindrical symmetry, also known as Vector Vortex Beams (VVB), are nowadays routinely used in fields such as laser material processing, optical tweezers, high-resolution microscopy, optical metrology, and classical and quantum communication, among many others \cite{Zhan2009,Woerdemann2013,Roadmap,Ndagano2017,Milione2015}. The generation of VVB has been achieved internal or external to laser cavities, using geometrical phase elements \cite{Niv2004,Marrucci2006,Naidoo2016} or optical interferometers \cite{Tidwell1990,Liu2012}. In the latter, two vortex beams with opposite topological charge and orthogonal polarizations are recombined interferometrically to generate VVB with various polarization distributions and spatial shapes. Remarkably, interferometric techniques have been fueled by the advent of Spatial Light Modulators (SLM), which have provided with one of the most flexible and versatile methods \cite{Forbes2016,SPIEbook,Maurer2007}. Crucially, none of the existing techniques has demonstrated the simultaneous generation of multiple VVB,  even though many research fields would greatly benefit from them. Importantly, the simultaneous generation of multiple scalar beams (multiplexing) is possible with the use of SLMs, which has found applications in optical tweezers and optical communications \cite{Cismar2010,Wang2012a,Bozinivic2013,Trichili2016,Li2016}.

Most common interferometric methods, aimed at  the generation of VVB, rely on splitting the amplitude of a scalar beam into two new beams to manipulate their polarizations, topological charge and amplitude independently. Their topological charge can be manipulated via SLMs or spiral phase plates while their polarizations can be controlled with half- and quarter-wave plates. VVB are then generated from a coaxial superposition of these beams carrying opposite topological charges and orthogonal polarization helicities.  Manipulation of the spatial shape and  polarization of each beam or a phase delay between them allows to switch between different vector beams. This approach, in general, can only produce one vector beam at a time. 

Here we put forward a novel generation technique that enables the rapid generation of any VVB and allows for the simultaneous generation of multiple vector beams (vector-beam multiplexing) using a single hologram. Our proposed method is also interferometric in nature but differs from the rest as it relies on dividing the wavefront of the initial beam and not its amplitude. This allow us to split the original beam into many that when recombined in pairs enables the generation of multiple vector beams simultaneously. Since this technique is purely digital, it does not require the manipulation of external optical components. That is, any combination of multiple vector beam can be generated by simply changing the digital hologram displayed on the SLM. As proof-of-principle, we demonstrated the simultaneous generation of twenty VVB, each with different spatial shapes and polarization distributions. Even though we restricted to the generation of VVB, our digital method enables the generation of vector beams with arbitrary polarization distributions and spatial symmetry.

\section{Concept}
VVB can be generated as linear combinations of optical vortices of opposite topological charge $\ell$ (with $\ell\in\mathbf{Z}$) and orthogonal polarization helicities as \cite{Milione2011,Galvez2012,Chen2014},  
\begin{equation}
|\Psi\rangle=\Psi_R^{-\ell}|-\ell\rangle|R\rangle\text{e}^{i\alpha_1}+\Psi_L^\ell|+\ell\rangle|L\rangle\text{e}^{i\alpha_2}
\label{VB}
\end{equation}
\begin{figure}[b]
\centering
\includegraphics[width=1\textwidth]{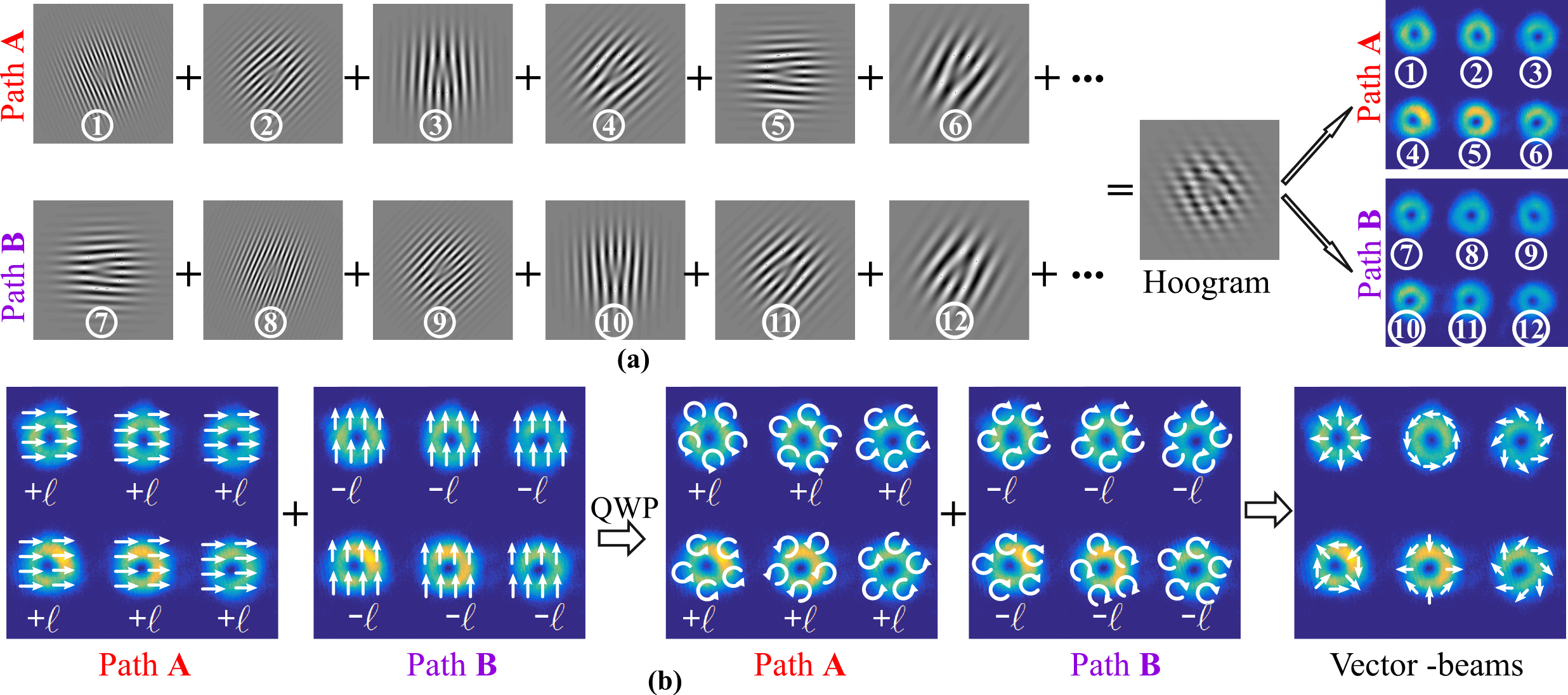}
\caption{(a) Several holograms are multiplexed with appropriate gratings to generate two sets of beams, one traveling along path {\bf A} and another along {\bf B}. (b) By recombining the beams on path {\bf A} and {\bf B} with appropriate orthogonal polarizations any vector beam can be generated.}
\label{Concept}
\end{figure} 
where, the vectors $|R\rangle$ and $|L\rangle$ represents the right and left circular polarization with corresponding amplitudes $\Psi_R^\ell$ and $\Psi_L^\ell$. Hence, in order to generate multiple vector beams simultaneously, two independent sets of scalar beams traveling along different paths have to be first generated. This can be easily achieved on the SLM by multiplexing each beam's hologram with different spatial gratings into a single hologram. The spatial shape, amplitude and phase of each beam can be manipulated accordingly via the digital hologram. The purpose of sending beams along different paths is to manipulate their state of polarization independently to achieve orthogonal polarization helicities $|L\rangle$ and $|R\rangle$. Coaxial superposition of these two sets of beams, carrying opposite topological charges and orthogonal polarization helicities, generates the desired vector beams. Manipulation of the spatial shape and phase difference between both beams, via the digital hologram, makes possible the generation of any vector beam on the High Order Poincar\'e Sphere (HOPS). Figure \ref{Concept} illustrates our generation principle. Two sets of holograms are multiplexed on a single hologram (Fig. \ref{Concept} (a)) to generate two sets of beams, one traveling along path {\bf A} and the other along path {\bf B} (\ref{Concept} (b)). The state of polarization of both paths is then manipulated so that the beam along one path  carries orthogonal helicity with respect to the beam along the other path. Afterwards, beams from one set are recombined with beams from the other set to generate multiple vector beams as shown in Fig. \ref{Concept} (b)). To illustrate the concept Fig. \ref{Concept} shows only six vector beams but as it has being shown before, it is possible to multiplex close to 200 scaler modes \cite{Rosales2017}, which will lead to the generation of one hundred vector modes.

\section{Experimental setup}
\begin{figure}[b]
\centering
\includegraphics[width=\textwidth]{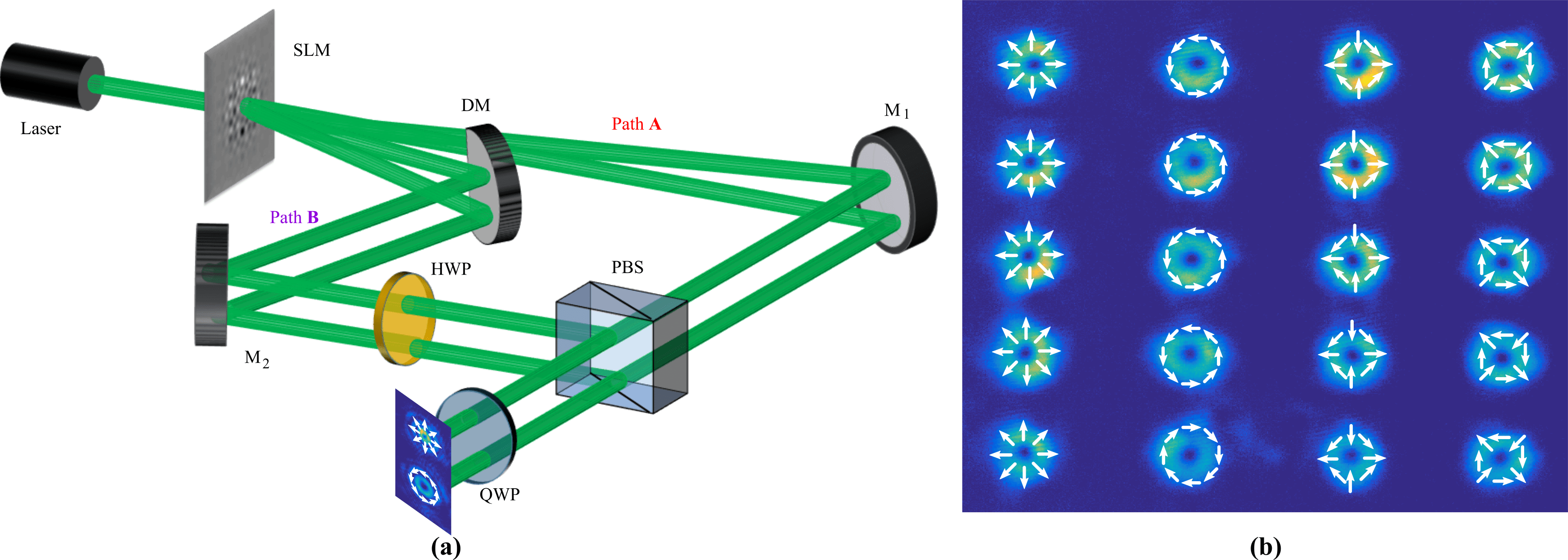}
\caption{(a) Schematic representation of the implemented setup to generate multiple vector beams. SLM: Spatial light modulator; DM: D-shaped mirror; M$_{1,2}$: mirrors; PBS: Polarizing beam splitter; QWP: Quarter-wave plate; HWP: Half-wave plate. (b) Multiplexing of twenty vector vortex beams with various polarization distributions.}
\label{setup}
\end{figure}

To experimentally demonstrate the simultaneous generation of multiple vector beams, we used a horizontally polarized laser  ($\lambda=532$ nm Verdi G Coherent) and a reflective SLM (Holoeye Pluto) to generate the scalar beams. Schematic representation of the implemented setup is shown in Fig. \ref{setup} (a).  On the SLM, the two independent sets of holograms to generates two sets of scalar fields are multiplexed, each beam with a unique grating. The grating periods are carefully chosen to separate the generated holograms into two groups, one traveling along path {\bf A} and the other along path {\bf B}. To accomplish this, the holograms are multiplexed such that the grating periods within each group are relatively small compared to the grating periods of the two different sets. The beams propagating along path {\bf A} maintains their horizontal polarization while the ones traveling along path {\bf B} are rotated to vertical polarization using a HWP at $45$$^{\circ}$. Both sets of beams are recombined into a single set, using a Polarising Beam Splitter (PBS). To simplify the overlapping of all beams, a judicious choice of the gratings is chosen so that the alignment of one ensures the alignment of the rest. A Quarter-Wave Plate (QWP) at $45$$^{\circ}$ transform their states of polarization to left- and right-handed, respectively, to generate in this way multiple VVBs, each with its own polarization state or spatial shape as illustrated in Fig. \ref{setup} (b). The resulting VVBs were observed with a CCD camera and their state of polarization analyzed with a linear polarizer placed before the CCD. In this proof-of-principle experiment we restricted ourselves to the generation of vector beams with cylindrical symmetry but any other vector beam can be generated by simply modifying the displayed hologram. 

\section{Results}
\subsection{Multiplexing of vector beams on the high order Poincar\'e sphere}
\begin{figure}[b]
\centering
\includegraphics[width=\textwidth]{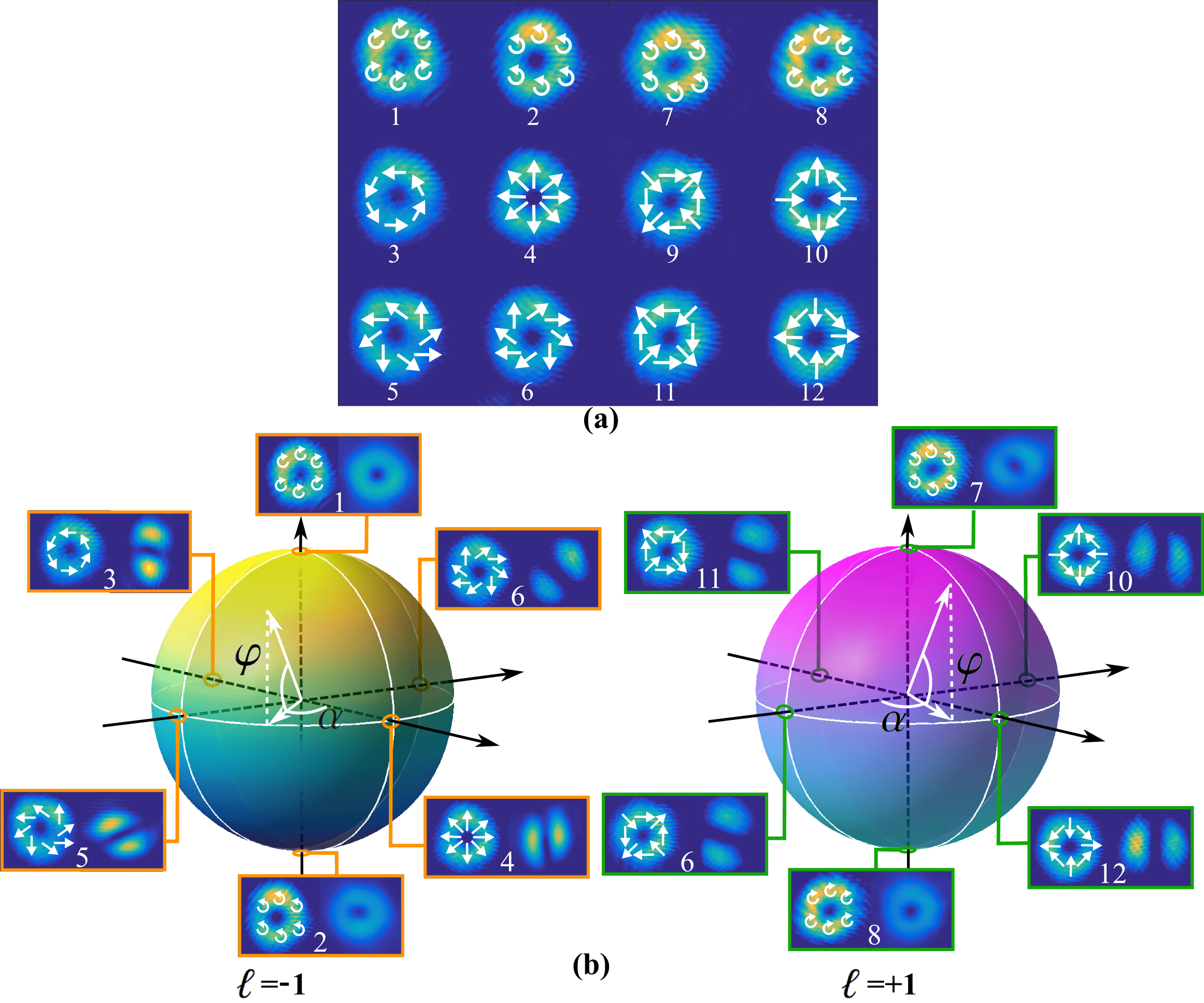}
\caption{(a) Intensity plot of twelve multiplexed vector beams from a single digital hologram. (b) Modes one to six correspond to a Poincar\'e sphere defined by a topological charge $\ell=-1$, whereas modes six to twelve corresponds to a topological charge $\ell=1$.}
\label{Pointcaresphere}
\end{figure}

The higher order Poincar\'e sphere (HOPS) is a very useful geometrical interpretation of cylindrical vector modes, according to which any vector state can be represented as a point ($\alpha$,$\varphi$) on a sphere \cite{Naidoo2016,Milione2011,Galvez2012}. In this representation, left and right circularly polarized states are positioned on the poles, all vector vortex beams are located along the equator and the remaining states occupies the rest of the sphere. A mathematical representation can be derived from Eq. \ref{VB} by choosing, $\Psi_R^{-\ell}= \cos(\varphi /2)/\sqrt{2}$, $\Psi_L^{\ell}= \sin(\varphi /2)/\sqrt{2}$ and $-\alpha_1=\alpha_2=\alpha/2$ as,

\begin{equation}
|\Psi_{\varphi,\alpha}\rangle=\frac{1}{\sqrt 2}\cos\left(\frac{\varphi}{2}\right)|-\ell\rangle|R\rangle{\text e}^{-i\alpha/2}+\frac{1}{\sqrt 2}\sin\left(\frac{\varphi}{2}\right)|\ell\rangle|L\rangle{\text e}^{i\alpha/2},
\end{equation}
where $\alpha \in [0, 2\pi]$ and $\varphi \in [0, \pi]$. In this representation, each $\ell$ value gives rise to a unique beam on the HOPS. The SLM-based method presented in this study enables the generation  of any vector mode on the full Poincar\'e sphere. Moreover, it allows for the generation of vector beams corresponding to different  HOPS. For example, by simply varying the phase offset $\text{e}^{i\alpha}$, we can generate any vector mode along the equator.  In addition, we can vary the weighting terms, related to $\varphi$, on either path of the interferometer  by changing the phase modulation depth of the encoded hologram to move from a full vector mode to a full scalar mode on either pole of the  HOPS \cite{Moreno2012,Chen2015a}. Fig. \ref{Pointcaresphere} illustrates the simultaneous generation of twelve states, six belonging to a HOPS given by $\ell=-1$ and six to the HOPS given by $\ell=+1$.  

\subsection{Multiplexing of Bell states}
As another example to highlight the capabilities of our technique, we multiplexed eight orthogonal Bell states, four of which were generated with $\ell=\pm1$ and the other four with $\ell=\pm2$. Bell states are of great relevance in optical communication and quantum computing \cite{Cox2016,Milione2015,Milione2015e}, mathematically they can be represented as, 
\begin{align}
|TM\rangle &= \frac{1}{\sqrt{2}}|\ell\rangle|R\rangle+|-\ell\rangle|L\rangle,\, &|HE^o\rangle &= \frac{1}{\sqrt{2}}(|\ell\rangle|L\rangle-|-\ell\rangle|R\rangle) \nonumber\\|TE\rangle &= \frac{1}{\sqrt{2}}(|\ell\rangle|R\rangle-|-\ell\rangle|L\rangle),\ & |HE^e\rangle & = \frac{1}{\sqrt{2}}(|\ell\rangle|L\rangle+|-\ell\rangle|R\rangle),
\label{Bell states}
\end{align}

\begin{figure}[t]
\centering
\includegraphics[width=.9\textwidth]{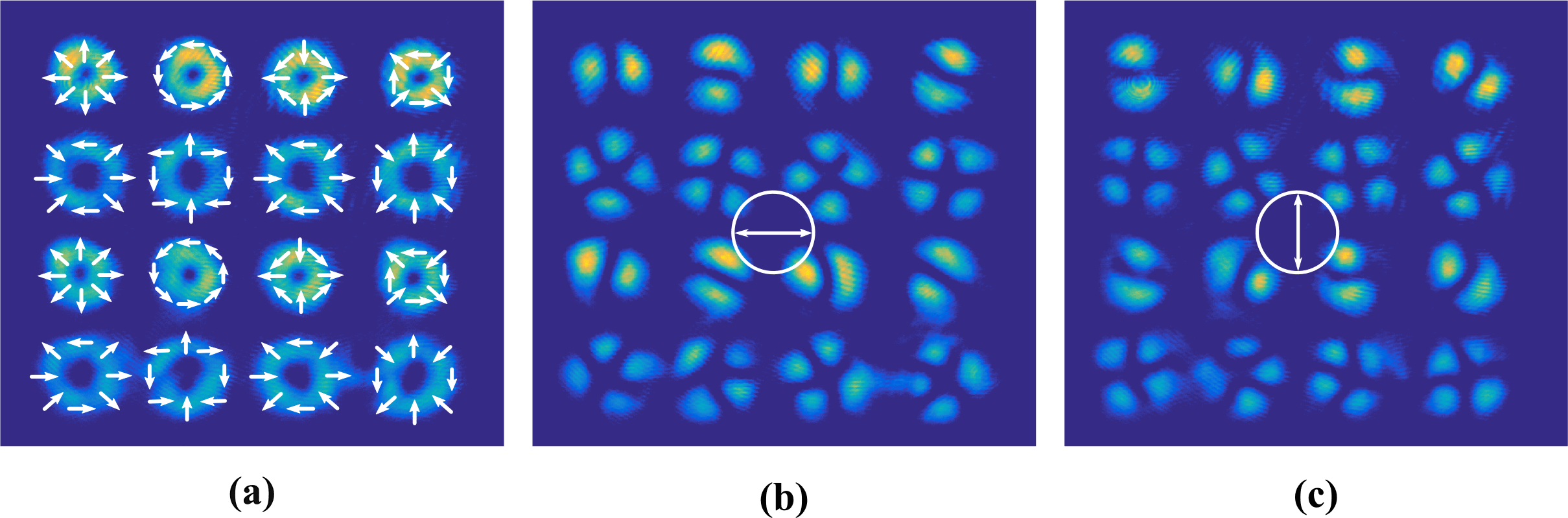}
\caption{(a) Experimental intensity profile of sixteen multiplexed VVB with eight orthogonal polarization states, indicated by the arrows, for $\ell=1$ and $\ell=2$. Intensity profile of the vector beams when a linear polarized is placed at  (b) $0^\circ$ and (c) $135^\circ$.}
\label{Cylindrical}
\end{figure}

Figure \ref{Cylindrical} shows the intensity profile of the generated Bell states, each of which was generated by properly adjusting the hologram's topological charge and phase. To generate the $|TM\rangle$ modes, two holograms with opposite topological charges $-\ell$ and $+\ell$ were multiplexed on the hologram, with the appropriate gratings to send each beam along path {\bf A} and {\bf B} respectively. A similar procedure is applied to generate the mode $|TE\rangle$, with the addition of a $\pi$ phase offset to one of the beams on the hologram. To generate the $|HE^o\rangle$ and $|HE^e\rangle$ modes, we simply interchange the $\ell$ values in the hologram. To demonstrate the power of our technique, we multiplexed a total of sixteen Bell states.   The intensity profile after a linear polarizer orientated at $\theta=0^\circ$ and $\theta=90^\circ$, inserted before the CCD, are shown in Figs. \ref{Cylindrical}(b) and  \ref{Cylindrical}(c) respectively. 

\subsection{Multiplexing of Vector Bessel beams}
\begin{figure}[t]
\centering
\includegraphics[width=\textwidth]{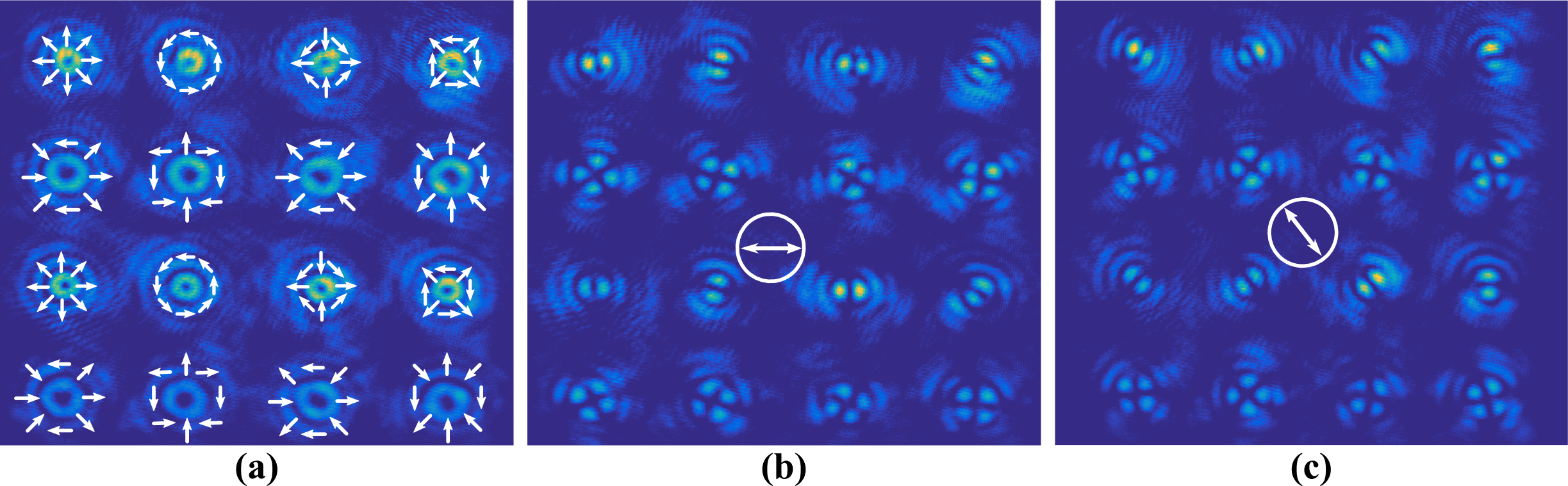}
\caption{(a) Experimental intensity profiles of sixteen multiplexed vector Bessel beams with eight orthogonal polarization states, indicated by the arrows, for $\ell=\pm1,\pm2$ and $\pm2$. Corresponding intensity profile when a linear polarized at angles (b) $0^\circ$ and (c) $135^\circ$ with the horizontal axis is placed.}
\label{BesselVector}
\end{figure}

To show that our technique can be applied to vector beams with other spatial shapes, we multiplexed sixteen Vector Bessel Beams (VBB). While previous methods have used SLMs to generate single VBB \cite{Avis2016,Dudley2013a}, here we the multiplexing of sixteen VBB simultaneously. To generate these modes, we can substitute the amplitude terms $\Psi_R^{-\ell}$ and $\Psi_L^{\ell}$ of Eq. \ref{VB} by $J_{-\ell}(k_t\rho)$ and $J_{\ell}(k_t\rho)$, respectively. Here, $J_\ell(x)$ is the Bessel functions of the first kind and $k_t$ is the transverse component of the wave vector. Bessel beams of any order can be easily generated by encoding a digital axicon an an SLM\cite{Rosales2017}. To generate multiple VBB simultaneously, we proceed in an analogous way as before by first generating two sets of scalar Bessel beams, $J_{\pm\ell}(k_t\rho)$. Experimentally, we can not generate a pure Bessel beam, as it will carry infinite energy. Instead, we generate a Bessel-Gaussian beam, by multiplying the term $J_{-\ell}(k_t\rho)$ by a Gaussian envelope $\exp(-\rho^2/\omega_0^2)$, where $\omega_0$ is the beam width. Figure \ref{BesselVector} (a) shows a set of sixteen multiplexed vector Bessel modes with $\ell=\pm1$, Figs. \ref{BesselVector} (b) and \ref{BesselVector} (c) shows the intensity profile when a linear polarizer at $\theta=0^\circ$ and $\theta=135^\circ$, respectively, is inserted in front of the CCD camera.     

\section{Discussion and conclusion} 
In this article we have presented a novel method to generate and multiplex vector modes using SLMs. Our method is based on the multiplexing concept enabled by SLMs, that allows to split an input light beam into many more, each with their own properties such as spatial shape and phase.  For our technique to work, two sets of multiple holograms are multiplexed with unique carrier frequencies on a single hologram. Each set of beams is sent along two spatially separated paths where they are endowed orthogonal helicities via phase retarders. In this scheme, the beams within each group are spatially separated to avoid overlapping among them. Both sets of beams are then coaxially superimposed to generate a full set of vector beams. The digital nature of this method allows full control of each beam's amplitude weighting, spatial shape and phase, which allows the generation of arbitrary vector beams with various spatial shapes and polarization distribution. Here we only generated cylindrical vector beams but our technique can be applied to any other geometry. Although we only generated 20 vector beams to demonstrate the concept, this technique allows for the generation of many more. Moreover, we believe that this generation technique also allows for the precise digital sorting of any vector beam, which is of great relevance in optical communications.

\section*{Funding}
National Research Foundation (NRF); the Claude Leon Foundation; CONACyT;CSIR-DST;  
\section*{Acknowledgement}
\end{document}